\documentclass[a4paper,10pt]{article}
\usepackage[utf8]{inputenc}
\usepackage{authblk}
\usepackage{fullpage}

\usepackage[section]{placeins}
\usepackage{graphicx}
\usepackage{epstopdf}
\usepackage{todonotes}
\usepackage{lineno,hyperref}
\usepackage{upgreek}
\usepackage{units}
\usepackage{multirow}
\usepackage{array}
\usepackage{amssymb}	
\usepackage{amsmath}	
\usepackage{textcomp}
\usepackage{pdfpages}
\usepackage{paracol}
\usepackage{wrapfig}
\usepackage{xcolor}

\usepackage{caption}
\usepackage{subcaption}

\title{Silicon Wafer Fracture Stress for Tracking Sensors in Particle Physics Experiments}

\author[1]{Haider Abidi}
\author[2]{Vitaliy Fadeyev}
\author[3]{Tim Jones}
\author[4, former 3]{Akhil Kumar}
\author[3]{Tom Lee}
\author[5,6]{Luise Poley}
\author[7]{Craig Sawyer}
\author[8]{Giorgio Vallone}
\author[5]{Sven Wonsak}

\affil[1]{Brookhaven National Laboratory, Rochester Street, Upton, United States of America}
\affil[2]{Santa Cruz Institute of Particle Physics, University of California, High Street, Santa Cruz, USA}
\affil[3]{Department of Physics, University of Liverpool, Oxford Street, Liverpool, United Kingdom}
\affil[4]{Institute of Astronomy at the University of Cambridge, Madingley Road, Cambridge, United Kingdom}
\affil[5]{TRIUMF, Wesbrook Mall, Vancouver, Canada}
\affil[6]{Department of Physics, Simon Fraser University, University Drive W, Burnaby, Canada}
\affil[7]{Particle Physics Department, STFC Rutherford Appleton Laboratory, Harwell Science and Innovation Campus, Didcot, United Kingdom}
\affil[8]{Lawrence Berkeley National Laboratory, Cyclotron Road, Berkeley, USA}

\begin{document}

\maketitle

\begin{abstract}

For the construction of the ATLAS Inner Tracker strip detector, silicon strip sensor modules are glued directly onto carbon fibre support structures using a soft silicone gel. During tests at temperatures below \unit[-35]{$^{\circ}$C}, several of the sensors were found to crack due to a mismatch in coefficients of thermal expansion between polyimide circuit boards with copper metal layers (glued onto the sensor) and the silicon sensor itself. While module assembly procedures were developed to minimise variations between modules, cold tests showed a wide range of temperatures at which supposedly comparable modules failed. The observed variance (fracture temperatures between \unit[-35]{\textcelsius} and \unit[-70]{\textcelsius}) for supposedly comparable modules suggests an undetected variation between modules suspected to be intrinsic to the silicon wafer itself. Therefore, a test programme was developed to investigate the fracture stress of representative sensor wafer cutoffs. This paper presents results for the fracture stress of silicon sensors used in detector modules.

\end{abstract}

\section{Introduction} 

Modules for the ATLAS~\cite{ATLAS} Inner Tracker (ITk) strip detector~\cite{ITk-TDR} are assembled by gluing kapton PCBs with a thickness of \unit[250 to 400]{$\upmu$m}, consisting of layers of polyimide, adhesive and copper, onto silicon strip sensors with a thickness of \unit[$320 \pm 15$]{$\upmu$m} with an epoxy glue~\cite{ABC130modules},~\cite{George}. Completed modules, loaded onto rigid mechanical support structures using a soft silicone gel, have been found to crack at temperatures below \unit[-35]{$^{\circ}$C}~\cite{Matt},~\cite{Cracking}. An analysis of the failure supported by mechanical simulation identified the mismatch in coefficient of thermal expansion (CTE) of the sensor and the PCBs glued onto them as the underlying cause~\cite{simref}. When the assembly is cooled down, the PCBs contract and bend, leading to a deformation of the attached silicon sensor. This deformation was found to result in cracks of the sensor if the local stress exceeded the sensor fracture stress.

Cold tests of nominally similar modules on the same support structure showed that despite the use of identical components, tools and assembly methods, the sensors on different modules cracked at a wide range of temperatures (between \unit[-35]{$^{\circ}$C} and \unit[-70]{$^{\circ}$C}). Simulating the known variations between modules (e.g. glue layer thickness and volume) was found to result in much smaller resulting stress variations than the effect from the difference of fracture temperatures.

For example, in a simple approximation where stress is proportional to $\Delta$T (the temperature difference between assembly temperature and temperature at fracture), the stress at \unit[-70]{$^{\circ}$C} would be higher than at \unit[-35]{$^{\circ}$C} by a factor of 1.64. In comparison, the fracture rate of sensors glued to carbon fibre support structures can increase from \unit[0]{\%} to over \unit[90]{\%}.

In order to investigate the observed discrepancy, sensors and cutoffs from multiple wafers were used for fracture tests to infer the inherent influence of the sensor material.

\section{Devices Under Test}

All the samples were manufactured by Hamamatsu Photonics as part of the ATLAS ITk strip sensor production~\cite{Nobu} consisting of 25,000 sensors total. The majority of results presented here are based on test structures produced on ATLAS18LS wafers. They feature a pure p-doped bulk for most of the thickness, with the top few microns being modified by patterning and passivation. This side contains strip electrodes to collect the mirror charges produced by traversing charged particles. The back side has a uniform implant metallised with a thin aluminium layer --- further details can be found in reference~\cite{Nobu}. All sensor devices undergo an extensive Quality Control (QC)~\cite{Paul} and Quality Assurance (QA)~\cite{Eric} verification programme.

\begin{figure}[htbp]
\centering
  \includegraphics[width=\textwidth]{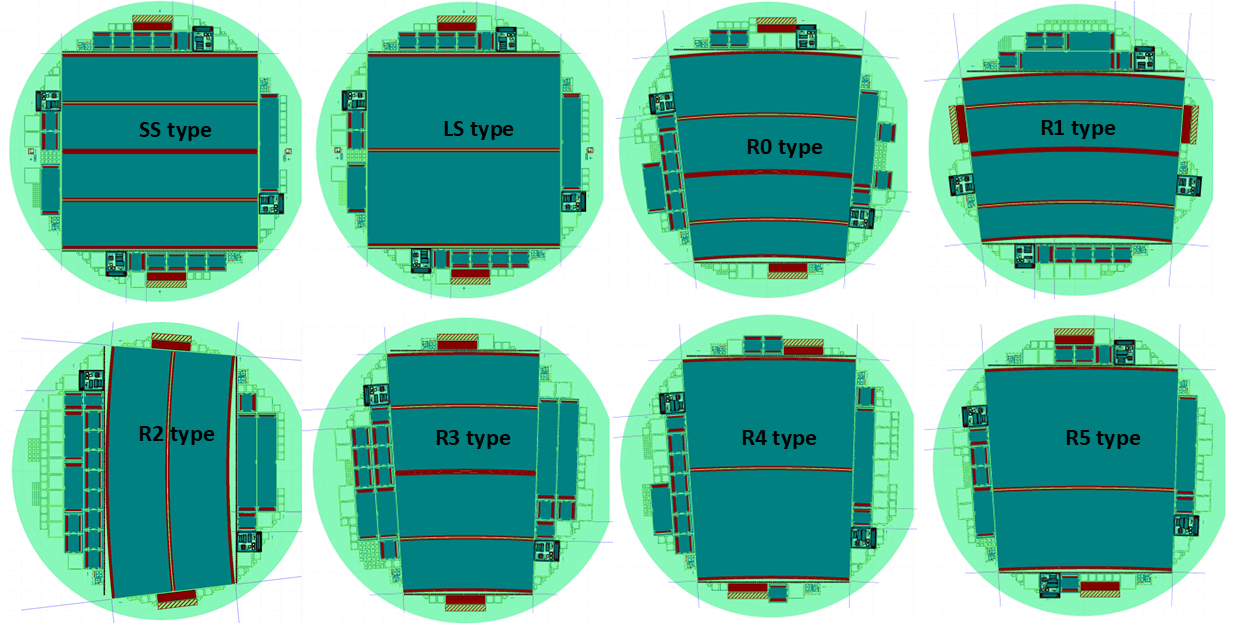}
  \caption{Wafer layouts for ATLAS ITk silicon strip sensors. The two square shapes at the top left are designed for the central part of the tracker (``barrel''). The other shapes were designed using different shapes and sizes for hermetic coverage in the forward part (``end-cap''~\cite{Carlos}). Depending on the geometry of the main sensor, the corresponding sensor halfmoon geometries show different shapes and sizes as well.}
  \label{fig:layouts}
\end{figure}

Tests were performed on two types of samples: full size sensors as used for detector modules as well as the cutoffs (called ``halfmoons'') from full sensor wafers after dicing out the main sensor shape. Figure~\ref{fig:layouts} shows an overview of the wafer layouts for different sensor geometries used in ITk strip tracker.

For comparison of cold test results obtained from sensors used for modules mounted on carbon fibre support structures, the corresponding halfmoons were selected for fracture stress tests.

In addition, full size sensors were chosen at random from the available inventory of sensors to compare the impact of the sensor geometry on the sensor fracture stress.

\section{Measurement Setup}
\label{setup}

\begin{figure}[htbp]
\centering
  \includegraphics[width=0.6\linewidth]{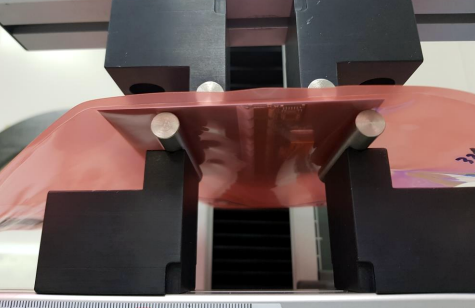}
  \caption{Setup of a Long Strip module in a four-point-bender, with the strip side in extension (facing the bottom rollers). This configuration was also used for all halfmoons large enough to fit the standard geometry.}
  \label{fig:setup}
\end{figure}

All measurements were performed in a four point bender setup (see figure~\ref{fig:setup}) using a Lloyd Instruments LRX Plus Universal Materials Tester.

Table~\ref{tab:setup} provides an overview of the positions of all rollers for the shapes of the various silicon pieces under test. Most of the end-cap halfmoons were large enough (width at the widest part above \unit[90]{mm}, referred to as ``Large halfmoons'') to be tested in the same configuration as barrel halfmoons and sensors. Smaller halfmoons (``Small halfmoons'') were tested in a special setup configuration where the four rollers were moved closer together.

\begin{table}[htp]
    \centering
    \begin{tabular}{c|cc}
         & Large halfmoons & Small halfmoons\\
         \hline
         Outer roller distance (centre-centre), $[$mm$]$ & 70 & 45 \\
         Inner roller distance (centre-centre), $[$mm$]$ & 30 & 15 \\
         \hline
         & \multicolumn{2}{c}{Common geometry} \\
         Roller length, $[$mm$]$ & \multicolumn{2}{c}{105} \\
         Roller diameter, $[$mm$]$ & \multicolumn{2}{c}{10} \\
         \hline
         Full Size sensors & X & \\
         Barrel halfmoons & X & \\
         Large end-cap halfmoons & X & \\
         Small end-cap halfmoons &  & X \\
    \end{tabular}
    \caption{Four-point-bender configurations for different fracture stress measurements. Identical settings were used for all tests except for end-cap sensor halfmoons which were too small for the default setup configuration.}
    \label{tab:setup}
\end{table}

Prior to each measurement, the setup was aligned to ensure that all rollers were parallel and at the same height compared to the sample under test using flat metal bars.

\begin{figure}[htbp]
    \centering
    \includegraphics[width=0.5\linewidth]{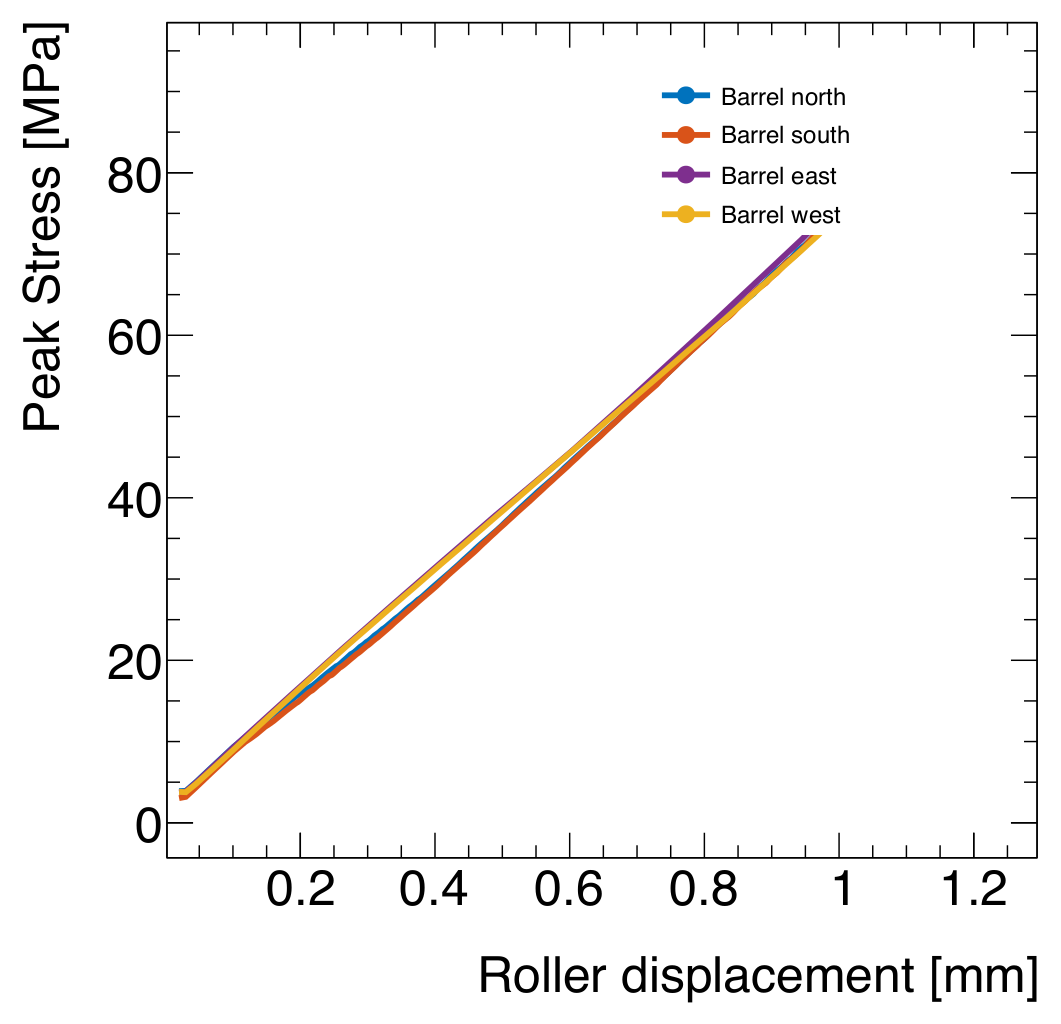}
    \caption{Simulation of the stress corresponding to a given displacement for different cutoff geometries. While the results are similar for different geometries, different geometries were accounted for to ensure a correct conversion from load to corresponding stress.}
    \label{fig:simulation}
\end{figure}

Since the different halfmoons from a wafer show minor variations in size and geometry (see section~\ref{subsec:HMgeo} for more details), an ANSYS~\cite{ansys} simulation of the material tester setup was set up to convert the measured load into a corresponding stress for each halfmoon. Figure~\ref{fig:simulation} shows the stress corresponding to a given displacement of the test setup.

Table~\ref{tab:original} shows the Weibull~\cite{Weibull} fit results of the peak load (referring to the load at fracture) for different measurement configurations: temperature (tests at \unit[+20 or -40]{$^{\circ}$C}) and sensor orientation. Within the four-point-bender setup, halfmoons were either placed so that the strip implant carrying side was facing towards the inner rollers of the setup (``face up''), meaning that, during the measurement, the sensor side with strip implants would be compressed, or with the strip implant side facing away from the inner rollers (``face down''), meaning the strip side of the sensor was extended during the measurement.
\begin{table}[htp]
    \centering
    \begin{tabular}{l|cc}
         & Peak load $[$N$]$, & Peak load $[$N$]$, \\
         & at \unit[+20]{$^{\circ}$C} & at \unit[-40]{$^{\circ}$C}\\
         \hline
        Strip side extended (``face down''))& 37.2$\pm$2.1 & 36.6$\pm$2.0 \\
        Strips side compressed (``face up'')& 21.9$\pm$0.6 & 20.6$\pm$0.6 \\
    \end{tabular}
    \caption{Weibull fit results for different configurations of halfmoon stress measurements.}
    \label{tab:original}
\end{table}
While the measurements at \unit[+20 and -40]{$^{\circ}$C} are mostly in agreement, a large difference was observed between the two cutoff orientations.

\section{Results}

\subsection{Initial Results}

\begin{figure}[htbp]
    \centering
    \includegraphics[width=\linewidth]{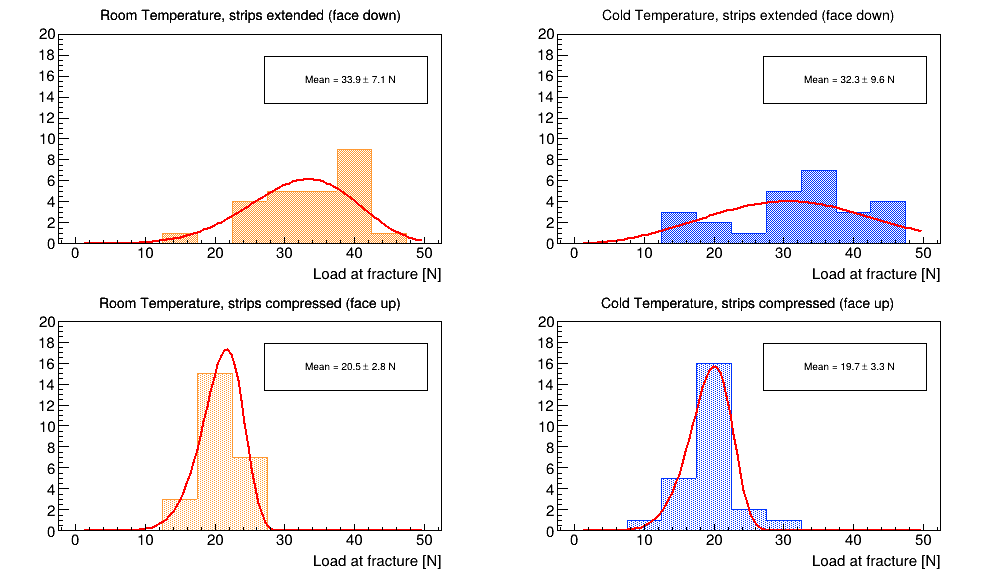}
    \caption{Fracture test results for halfmoons from randomly selected wafers at different temperatures and for different orientations. While the temperature was found to have only a minor impact on the fracture stress distribution, the orientation of the test pieces was found to affect the results significantly.}
    \label{fig:original}
\end{figure}

Initial tests in the four point bender setup showed a wide distribution of stress observed at fracture for the halfmoons tested ``face down'' (see figure~\ref{fig:original}). 

However, the measurements were performed on 100 cutoffs from randomly selected wafers (25 per configuration). They therefore did not provide information regarding how the measured fracture stress related to the reliability of a full sensor or module.

Therefore, an additional set of measurements was performed on silicon pieces from the same wafers used to assemble modules that had been subjected to cold tests, providing an opportunity to correlate the inherent fracture stress of parts from the same silicon wafers.

\subsection{Halfmoon geometries}
\label{subsec:HMgeo}

\begin{figure}[htbp]
    \centering
    \includegraphics[width=0.6\linewidth]{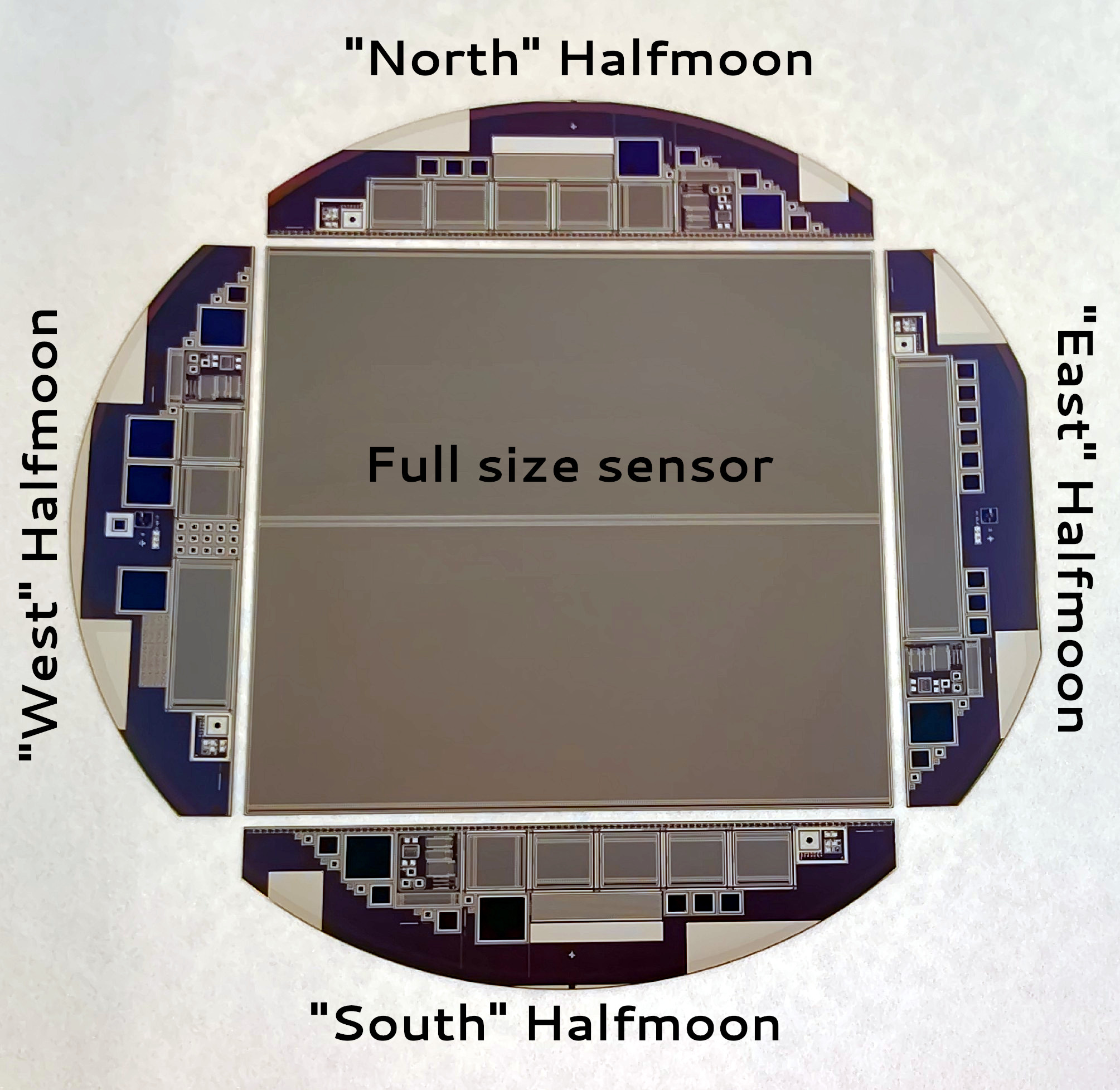}
    \caption{Long Strip sensor wafer layout with the main sensor at the centre and four cutoffs (``halfmoons'') surrounding it. The ``East'' halfmoon is located in an area of the wafer where the crystal lattice orientation is indicated by a straight cut on the wafer, leading to two long parallel edges on this halfmoon.}
    \label{fig:LS-Wafer}
\end{figure}

An initial measurement was set up to compare the fracture stress for different geometries of sensor halfmoons. Figure~\ref{fig:LS-Wafer} shows a detailed view of a Long Strip sensor wafer with one main sensor and four surrounding halfmoons, labeled as ``North'', ``East'', ``South'' and ``West'' for orientation.

\begin{figure}[htbp]
    \centering
    \includegraphics[width=\linewidth]{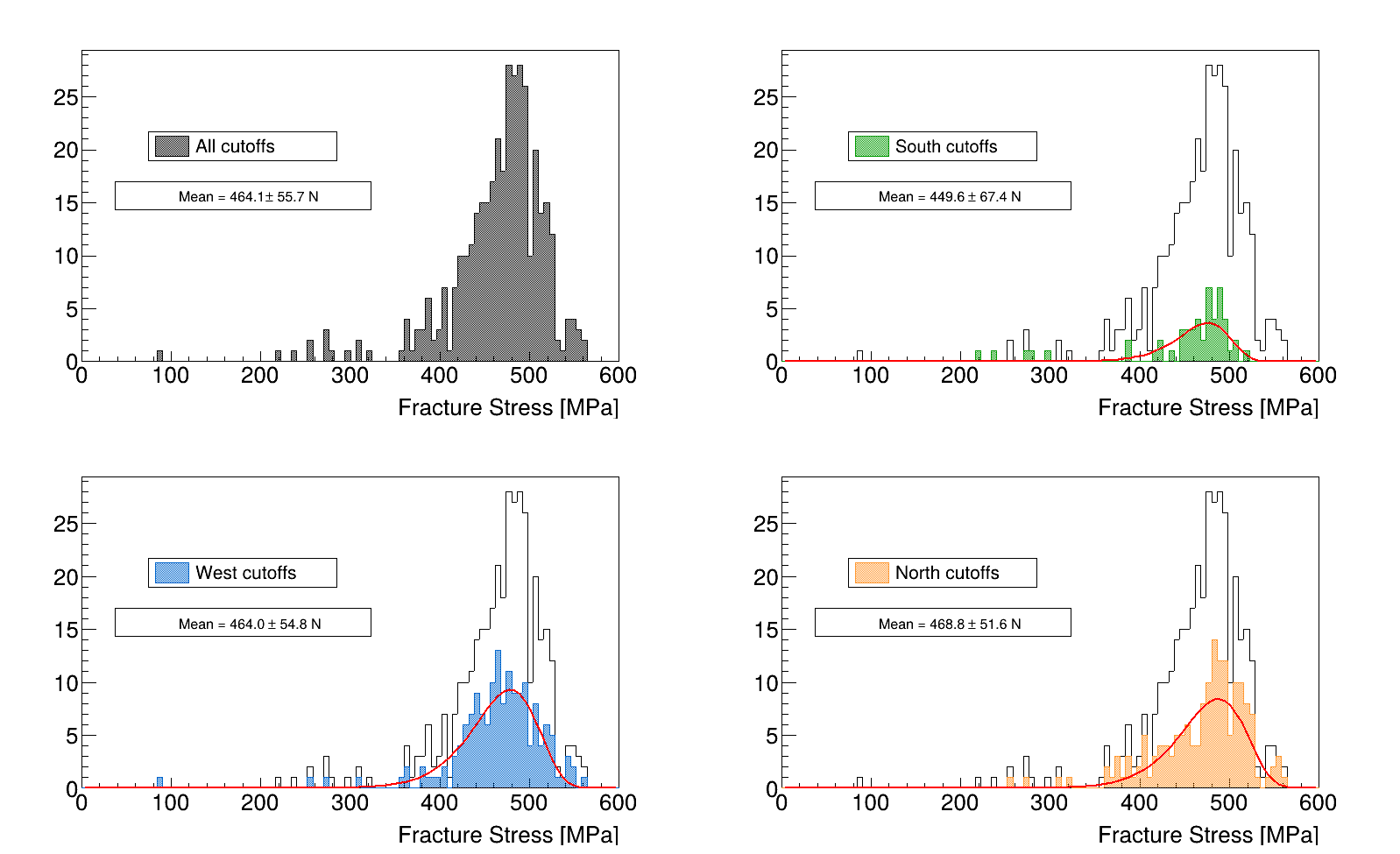}
    \caption{Fracture stress distribution for different geometries of Long Strip sensor wafer halfmoons. All halfmoons were tested with the top side in extension. The top left figure, added to the other plots for comparison, shows the combined distribution for all halfmoons. The red lines indicate Weibull fits of each distribution to determine each distribution's peak.}
    \label{fig:halfmoon-results}
\end{figure}

The locations of all halfmoons on each wafer was tracked throughout the measurement. Loads were converted to a corresponding stress accounting for each halfmoon shape.
Figure~\ref{fig:halfmoon-results} show the resulting peak stress distributions from a total of 380 halfmoons.

The position of each distribution's peak was determined from a Weibull fit. The determined peak locations are shown in table~\ref{tab:HM_peaks}.

\begin{table}[htp]
    \centering
    \begin{tabular}{lr}
        Geometry & Peak location, $[$MPa$]$\\
        \hline
        North & 490.2 $\pm$ 3.5\\
        West & 481.3 $\pm$ 3.3\\
        South & 477.7 $\pm$ 6.0\\
        \hline
        Combined & 486.3 $\pm$ 1.8\\
    \end{tabular}
    \caption{Overview of peak fracture stress for different halfmoon geometries. ``East'' halfmoons were sent to the company for testing, so that halfmoons from the same wafers would be compared.}
    \label{tab:HM_peaks}
\end{table}

The distributions show an overall good agreement with deviations between the types, with the largest difference amounting to \unit[2.7]{\%}. This difference is consistent with literature values~\cite{two_percent}, where similar variations have been attributed to differences in the surface structures, which likely contributes to the values measured for the different halfmoon geometries as well.

\subsection{Halfmoon Orientation}

\begin{figure}[htbp]
    \centering
    \includegraphics[width=0.8\linewidth]{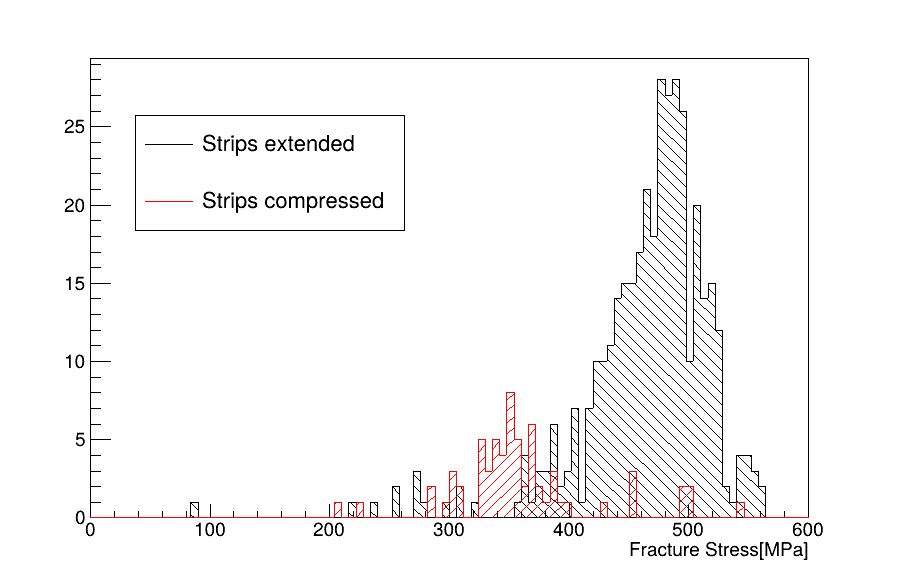}
    \caption{Peak stress results for wafer cutoffs tested with strips in extension and strips in compression.}
    \label{fig:up_vs_down}
\end{figure}

A subset of the tested halfmoons was used to investigate the effect of the halfmoon orientation in the bender machine on the fracture stress. Figure~\ref{fig:up_vs_down} shows the resulting fracture stress distributions for both orientations.

The obtained distributions showed a much lower fracture stress for sensor strip (top) sides in compression compared to an orientation where the strips were in extension: \unit[349$\pm$6 vs 486$\pm$2]{MPa}. The observed difference agrees with the direction of the original measurements (see figure~\ref{fig:original}) and matches expectations.

The fracture stress of a solid body depends on its defect size~\cite{strength}. During the semiconductor processing, the top surface undergoes treatments which remove deep sub-micron defects in order to increase the device yield. The top surface can therefore be expected to have a low defect density, corresponding to a high fracture stress.
The back of the wafer undergoes less stringent processing and can therefore be assumed to have a higher defect density.

The compression strength of silicon is larger than its tensile strength. Therefore, an orientation where the weaker sensor side (the back) is in compression and the side with fewer defects (the surface) is in extension can be expected to show a higher fracture strength.

This expectation agrees well with the obtained results. It was found that the stronger orientation (strip implants in extension) corresponded to the deformation direction on a cooled module. Therefore, this orientation was used for the majority of subsequent tests.

\subsection{Homogeneity within each wafer}

\begin{figure}[htbp]
    \centering
    \includegraphics[width=\linewidth]{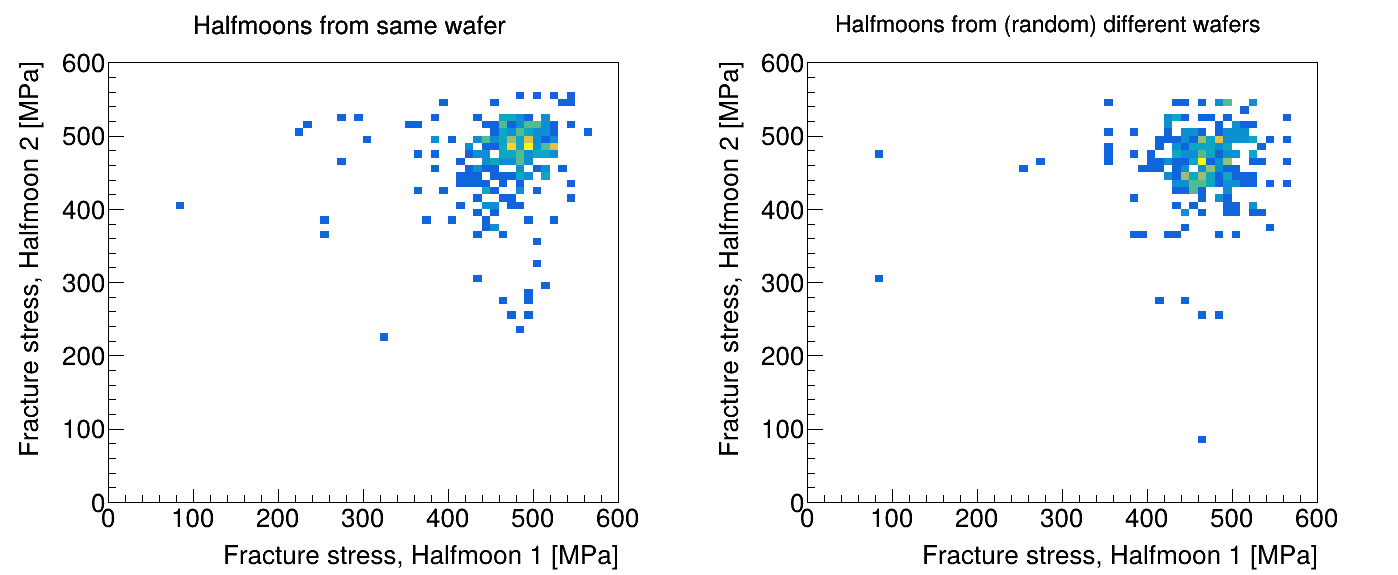}
    \caption{Investigating correlations between fracture stress results from the same wafer (left) compared with randomly selected pairs of halfmoons (right), excluding multiple halfmoons from the same wafer.}
    \label{fig:Corr_Marcel}
\end{figure}

Testing multiple halfmoons from the same wafer provided the opportunity to compare their individual fracture stresses. Figure~\ref{fig:Corr_Marcel} shows the correlation of fracture stress results measured for halfmoons from the same wafer (correlation factor 0.14) in comparison with pairs of halfmoons from randomly selected wafers (correlation factor 0.10). The resulting correlation plots shows that the fracture stress measured for a single halfmoon was not found to be a reliable indicator of the fracture stress for other objects from the same wafer.

\begin{figure}[htbp]
    \centering
    \includegraphics[width=\linewidth]{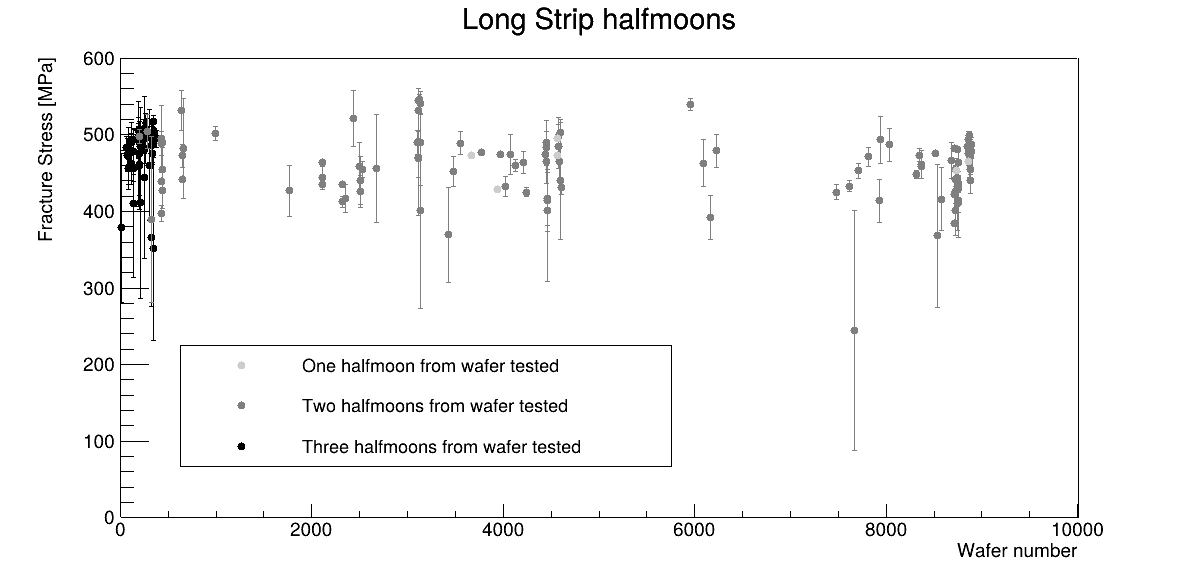}
    \caption{Mean fracture stress and spread (variance) per wafer, calculated from the measurements of the available number of cutoffs from the same wafer. As some of the halfmoons were used for other purposes, e.g. wire bonding trials or sensor QA tests, differing numbers of cutoffs per wafer were available for analysis.}
    \label{fig:LS_production}
\end{figure}

In addition to an analysis per wafer, the data was also investigated for potential trends over the course of production. Where available, multiple halfmoon test results per wafer were used to calculate the mean fracture stress per wafer and the spread for its parts. 
Figure~\ref{fig:LS_production} shows the resulting mean fracture stress and - where multiple results were available - spread for all tested halfmoons.

Despite spanning several years of sensor production, the obtained results showed consistent results over time. 
The results show a random distribution of consistency throughout the wafer - while a majority of wafers showed consistent results for several halfmoons, a significant fraction shows large variations.

Low fracture stress results on a single halfmoon were generally not confirmed by the other halfmoons from the same wafer. The defect distribution can therefore be assumed to be inhomogeneous throughout the wafer.

\subsection{Comparison of measurements performed in different setups}


In order to test the reliability of the results, another set of samples was tested at a commercial facility. For this test, only ``East'' halfmoons with two parallel edges were used in four-point-bender tests (see figure~\ref{fig:LS-Wafer}: halfmoon on the right side of the sensor).

\begin{figure}[htbp]
    \centering
    \includegraphics[width=\linewidth]{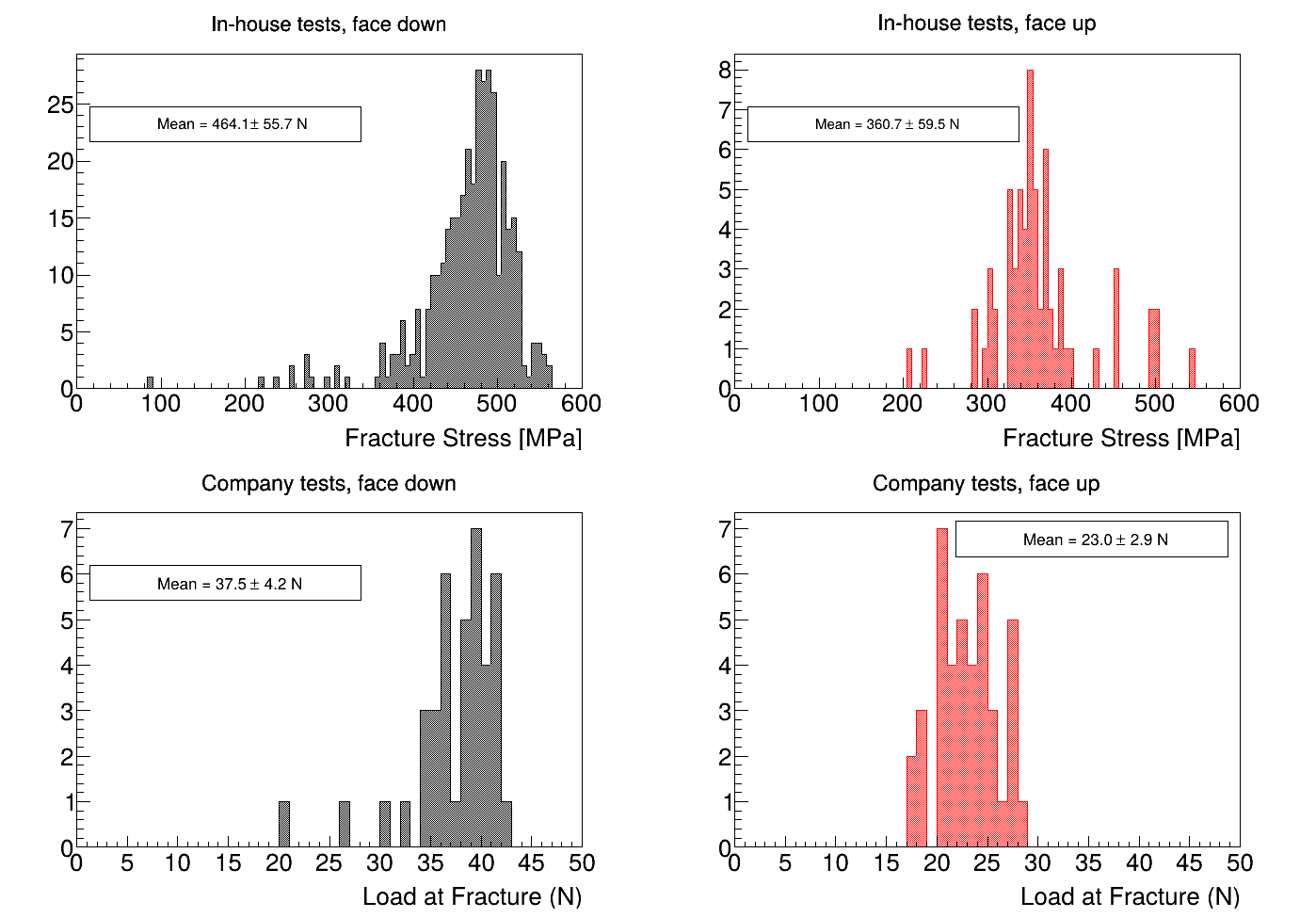}
    \caption{Fracture stress and load measured in two different setups for two halfmoon orientations facing up and down (top left distribution from figure~\ref{fig:up_vs_down}). The top left plot shows the combined entries for all geometries. It is reproduced in the other plots to facilitate the comparison between the types.}
    \label{fig:HM_Company}
\end{figure}

Figure~\ref{fig:HM_Company} shows a comparison of the measurements in two different locations. The measurements at the commercial facility were performed on a ZWICK Z050 (ProLine) material tester. The exact measurement configuration was unknown, hence the load at fracture could not be converted into a fracture stress reliably.

Comparing the fitted results from the different measurement configurations (see table~\ref{tab:company}) shows a significant difference between tests results for halfmoons in extension and compression. The differences between both orientations is roughly compatible for both setups.

\begin{table}[htp]
    \centering
    \begin{tabular}{c|cc}
         & Company test results, & In-house test results, \\
         & Load at Fracture $[$N$]$ & Fracture Stress $[$MPa$]$ \\
         \hline
        Top side in compression & $22.8 \pm 6.5$ &  $349 \pm 11$ \\
        Top side in extension & $39.2 \pm 9.8$ & $486 \pm 16$  \\
        \hline
        Ratio compression/extension & \unit[58$\pm$22]{\%} & \unit[72$\pm$2]{\%} \\
    \end{tabular}
    \caption{Summary of test results performed in two different setups. Note that the Company test results could not be converted to stress and are therefore given in units of applied load at fracture (N).}
    \label{tab:company}
\end{table}

It should be noted that this difference varied between both setups - tests at the commercial facility found a larger difference between test results in compression and extension than the in-house measurements (see table~\ref{tab:company}). 

\subsection{Correlation of fracture stress and sensor properties}

In order to investigate whether a lower fracture stress could be predicted by known sensor properties determined as part of non-destructive quality control.
They were obtained by ATLAS institutes performing QC tests after reception as well as from manufacturer test data. 

The fracture stress results were compared to the following mechanical and electrical test parameters as described in~\cite{Paul}:
\begin{itemize}
    \item Physical thickness
    \item Sensor bow
    \item Full depletion voltage as measured by manufacturer
    \item Full depletion voltage as measured upon reception
    \item Effective doping concentration
    \item Active thickness
    \item Leakage current at \unit[500]{V} as measured by manufacturer
    \item Leakage current per area, measured at \unit[500]{V} by ATLAS institutes
    \item Breakdown voltage
    \item Leakage current variance in a long-term stability test
    \item Percentage of failed sensor strips
\end{itemize}

\begin{figure}[htbp]
    \centering
    \includegraphics[width=0.7\linewidth]{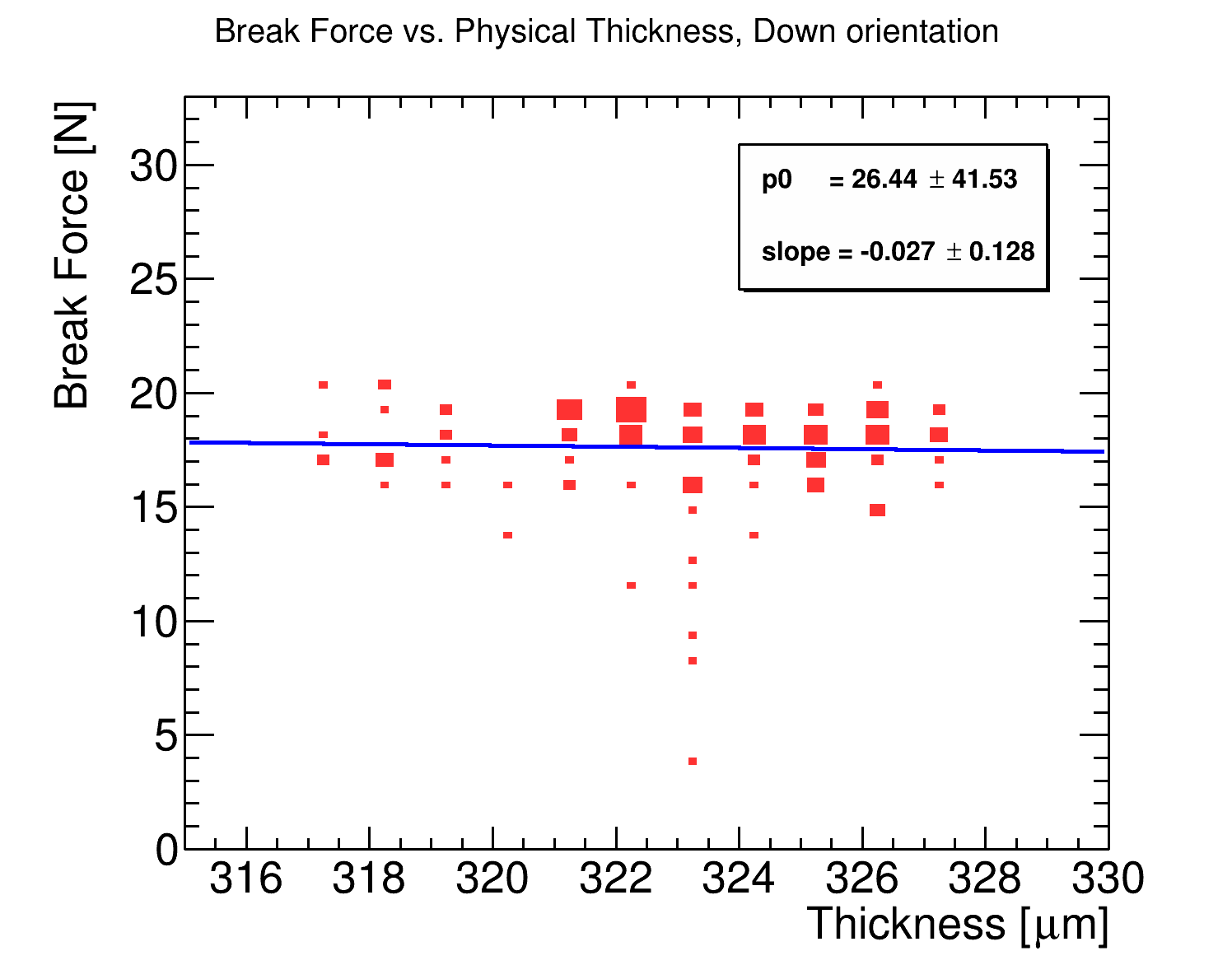}
    \caption{Correlation of the load at fracture with the physical sensor thickness. A linear fit is shown in the plot.}
    \label{fig:corr_TH}
\end{figure}

\begin{figure}[htbp]
    \centering
    \includegraphics[width=0.7\linewidth]{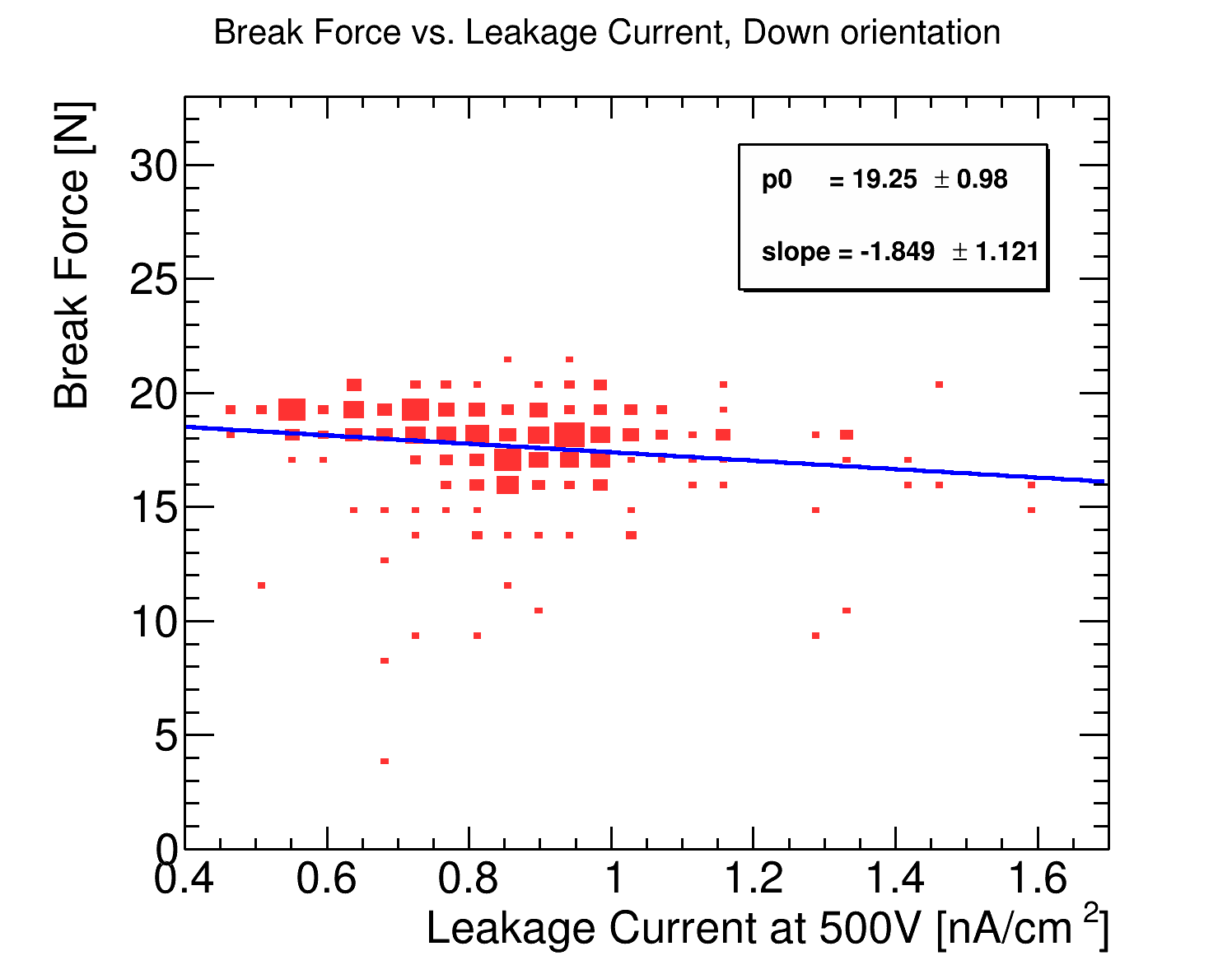}
    \caption{Correlation of the load at fracture with the measured leakage current at a reverse bias voltage of {\unit[500]{V}}. The linear fit parameters are shown in the plot.}
    \label{fig:corr_IV}
\end{figure}

Examples for two correlation plots are shown in figures~\ref{fig:corr_TH} and~\ref{fig:corr_IV}. 

In order to check for the presence of the correlations, each plot was fitted using a linear function. The fit slope for all parameters was found to be consistent with zero within the errors, implying no correlation with the load at fracture. The parameters with the most significant slope, at a \unit[1.65]{$\upsigma$} level, are the leakage current at \unit[500]{V} measured by ATLAS institutes and current variance in the stability test. The implication of these results is that sensors can not be pre-selected for higher stress tolerance on the basis of their known electrical or mechanical properties.

\subsection{Comparison of wafer halfmoons and full-size sensors}

\begin{figure}[htbp]
    \centering
    \includegraphics[width=0.8\linewidth]{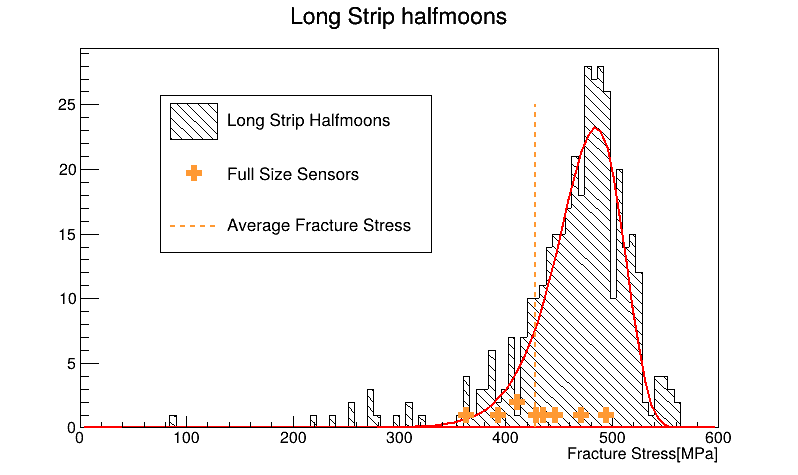}
    \caption{Comparison of the fracture stress measured for halfmoons vs full size sensors.}
    \label{fig:full_size}
\end{figure}

An important question for the underlying investigation was the comparability of halfmoon fracture test results and the corresponding fracture stress of full size sensors. Therefore, nine full size Long Strip sensors were used for fracture tests.
Figure~\ref{fig:full_size} shows a comparison between the fracture stress measured on individual halfmoons and full-size sensors.

Additionally, fully assembled modules were tested in the same setup to assess the fracture stress of a fully assembled module (see figure~\ref{fig:setup}).

A comparison of all results is shown in table~\ref{tab:large}. All objects were tested in an orientation where the strips were extended.
\begin{table}[htp]
    \centering
    \begin{tabular}{lr}
         & Fracture Stress, \\
         & $[$MPa$]$\\
         \hline
        Halfmoons (Weibull fit)& 486 $\pm$ 2\\
        Full size sensors (mean)& 428 $\pm$ 41\\
        Module 1 & 463 \\
        Module 2 & 464\\
    \end{tabular}
    \caption{Overview of measured fracture stresses and uncertainties for halfmoons (set of 380 samples), sensors (set of 7) and assembled modules. The fracture stresses measured for larger objects are consistent with the halfmoon results.}
    \label{tab:large}
\end{table}

Due to their larger area and therefore higher chance of localised defects coinciding with a high stress region, larger objects were expected to have a lower fracture stress than the smaller cutoffs. The measurements confirmed that larger objects had a lower fracture stress, but were overall similar to cutoffs, as expected for objects using the same material and similar surface structures. The module stress results are consistent with the full sensor results, within the statistical uncertainty.

\subsection{Other geometries}

\begin{figure}[htbp]
    \centering
    \includegraphics[width=\linewidth]{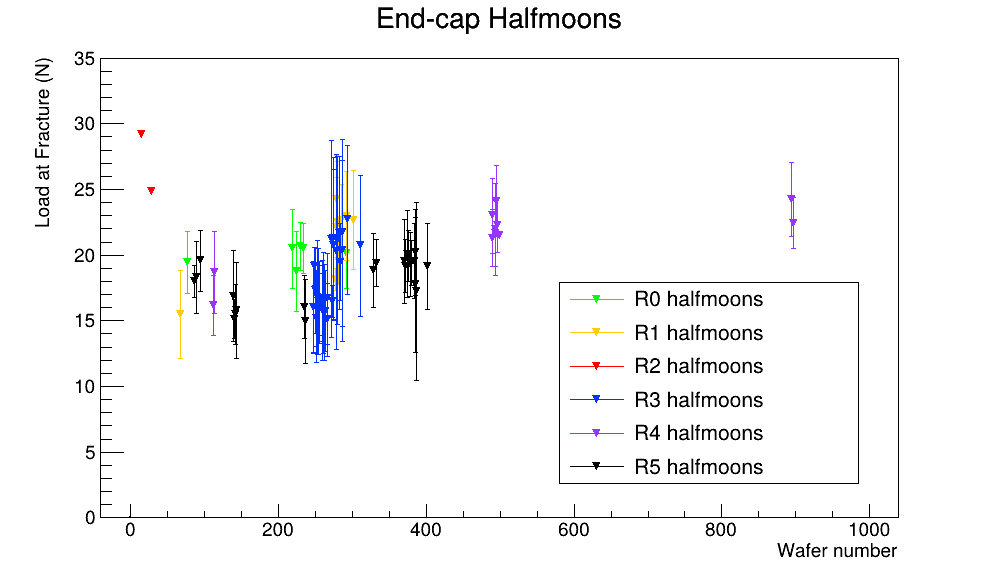}
    \caption{Halfmoon fracture load results and spread (variance) obtained from different halfmoon geometries, all shalfmoons were tested in ``face down'' orientation (strips in extension). Only individual R2 halfmoons per wafer were used in this test, therefore no spread could be calculated.}
    \label{fig:endcap}
\end{figure}

In addition to the halfmoons from barrel sensor wafers, where all cutoffs show similar shapes, a selection of cutoffs from other wafer geometries (see figure~\ref{fig:layouts}) was tested. Figure~\ref{fig:endcap} shows an overview of the results from a variety of halfmoon geometries.

It should be noted that the setup had to be adjusted to accommodate the very short halfmoons. For smaller pieces, the positions for the Four-Point-Bender setup were adjusted to ensure they would support each halfmoon at a width that corresponded to at least \unit[60]{\%} of its widest area.

Due to the small statistics, no gain was anticipated from converting the individual values into fracture stresses. 

All wafer types and shapes were consistent with the higher statistics of Long Strip Sensor halfmoons - loads at fracture were roughly comparable for all geometries.

\FloatBarrier

\section{Conclusion}

This paper investigated the fracture stress of a sample of silicon strip sensor cutoffs to assess their distribution, consistency over the course of production and predictability.

The tests found that the fracture stress measured on wafer cutoffs was consistent across about 400 Long Strip samples chosen from the first half of the full sensor production (25,000 wafers once completed). Samples showed a small dependence on the shape of the geometry of the wafer cutoff as well as a significant dependence on the its orientation, where testing the sensor top (strip) side in extension showed a significantly higher fracture stress than testing the top (strip) side in compression.

This outcome was found beneficial for the observed cracking of modules mounted on support structures, where the peak stress from CTE mismatches corresponds to the locations where the top sensor side is in extension. In this orientation the fracture stress distribution peaks at around \unit[490]{MPa}. However, there is a long low-side tail reaching less than half of this value. The special breaking tests performed at \unit[-40]{$^{\circ}$C} yielded results compatible with the ones at room temperature. This suggests that the large temperature difference by itself does not alter the wafer's mechanical properties.

Measurements of cutoffs from the same wafer showed large discrepancies in several cases, confirming that tests of individual cutoffs were not a reliable indicator of the fracture stress of the overall wafer or sensor. Comparing the measured fracture stresses of individual halfmoons to several electrical and mechanical properties of the corresponding sensor did not show any reliable correlations between measured properties and fracture stress.

Measurements of full size sensors showed about \unit[12]{\%} lower average fracture stress than measured for the smaller cutoffs.

\section*{Acknowledgements}

This work has been supported by the Canada Foundation for Innovation, the Natural Sciences and Engineering Research Council of Canada. The work at SCIPP was supported by the US Department of Energy, grant DE-SC0010107. The authors would like to thank the CERN-EN-MME, especially Oscar Sacristan De Frutos and Ana Teresa Perez Fontenla, for their help and support.

\bibliographystyle{unsrt}
\bibliography{bibliography.bib}

\end{document}